\documentclass[twocolumn,showpacs,preprintnumbers,amsmath,amssymb]{revtex4}

\usepackage{graphicx}
\usepackage{dcolumn}
\usepackage{bm}

\begin{document}

\preprint{}

\title{R-parity violating supersymmetric contributions to the \\
P, CP-odd electron-nucleon interaction at the one-loop level}

\author{Nodoka Yamanaka}
\affiliation{%
Department of Physics, Osaka University, Toyonaka, Osaka 560-0043, Japan}%

\date{\today}

\begin{abstract}
The contribution of the R-parity violating minimal supersymmetric Standard model (RPVMSSM) at the one-loop level to the $^{199}$Hg atomic electric dipole moment (EDM) through P, CP-odd electron-nucleon ($e-N$) interaction is calculated.
We show that the current experimental data of the $^{199}$Hg EDM give tighter constraints on some of the imaginary parts of R-parity violating (RPV) coupling than those currently known.
We add also the analysis of the P, CP-odd 4-quark interaction generated by R-parity violating interactions at the one-loop level, and discuss the possibility to constrain them in future experiments.
\end{abstract}

\maketitle

\section{\label{sec:intro}Introduction}
The Standard model (SM) of particle physics is known to be very successful in interpreting many experimental data up to now. There are however some phenomena which are difficult to explain in this framework, such as the matter abundance of our Universe. We need therefore extend the SM with some new physics (NP).

One approach to search for NP beyond the SM is the low energy approach, which consists of observing the small discrepancies between the measured low energy observables and the SM predictions. Among several others, the EDM experiments are of particular interest for the following reasons. The EDM is an observable sensitive to the violation of parity and time-reversal (or equivalently CP). The contribution from the SM is in general very small \cite{smedm}, but is sensitive to many NP with large CP violation. The experimental data available are very accurate for a variety of systems such as the neutron \cite{baker}, $^{205}$Tl atom \cite{regan}, $^{199}$Hg atom \cite{griffith} atoms, YbF molecule \cite{hudson}, and muon \cite{muong2}, which make the EDM to be a very efficient probe of NP.
Also new generation of experiment using storage ring is under preparation, aiming at the measurements of the EDMs of muon, proton and deuteron \cite{storage}.

On the theoretical side, the minimal supersymmetric Standard model (MSSM) is known to be the leading candidate of the NP. A general supersymmetric extension of the SM allows baryon number or lepton  number violating interactions, so we must impose the conservation of {\it R-parity} ($R=(-1)^{3B-L +2s}$) to forbid them. This assumption is however completely {\it ad hoc}, so the R-parity violating (RPV) interactions have to be investigated phenomenologically. Until now, many of the RPV interactions were constrained by high energy experiments, low energy precision tests, and cosmological phenomenology \cite{chemtob, barbier,rpvphenomenology}.
Thanks to many efforts in EDM experiments, many phenomenological analyses of the supersymmetry with \cite{susyedm1-loop,susyedm2-loop,susyedmgeneral,pospelovreview,susyedmflavorchange} and without \cite{barbieri,godbole,chang,herczeg,choi,faessler,yamanaka1} R-parity were done, and many CP phases have been constrained so far.
Herczeg analyzed the contribution of the RPVMSSM to the P, CP-odd $e-N$ interactions, and gave new constraints from the atomic EDM on the imaginary part of many combinations of RPV couplings at the tree level \cite{herczeg}. The P, CP-odd $e-N$ interactions receive a stringent constraint from the experimental data of atomic EDM \cite{khriplovich1,ginges,pospelovreview}. 
The present tightest limits on the P, CP-odd $e-N$ interactions are given by the recent update of the $^{199}$Hg EDM experiment as \cite{griffith}
\begin{equation}
d_{\rm Hg} < 3.1 \times 10^{-29} e \, {\rm cm} \ ,
\label{eq:hgedmlimit}
\end{equation}
The accuracy of the $^{199}$Hg EDM data is such that one can expect to constrain RPV parameters even at the one-loop level. This is because loop level diagrams can involve new RPV structures not encountered at the tree level.
The main purpose of this paper is then to investigate the P, CP-odd $e-N$ interaction at the one-loop level within RPVMSSM. Our discussion is organized as follows. In the next section, we briefly present the RPV interactions, and list the contributing diagrams to the P, CP-odd $e-N$ interaction at the quark level.
We then derive the EDM of $^{199}$Hg atom from this one-loop contribution.
In Sec. \ref{sec:analysis}, we analyze the bounds on RPV couplings that can be set from the $^{199}$Hg EDM data.
We also add the analysis of the P, CP-odd 4-quark interaction within RPVMSSM at the one-loop level and discuss the possibility to constrain RPV interactions with future experiments.
We finally summarize our discussion.

\section{\label{sec:RPV}RPV contribution}
In the first step of the estimation, we construct the one-loop level contribution to the P, CP-odd $e-N$ interaction at the quark level within RPVMSSM. The superpotential of the RPV interactions can be written as follows:
\begin{eqnarray}
W_{{\rm R}\hspace{-.5em}/} &=& \frac{1}{2} \lambda_{ijk} \epsilon_{ab}
 L_i^a L_j^b (E^c)_k
 +\lambda'_{ijk} \epsilon_{ab} L_i^a Q_j^b ( D^c)_k \nonumber\\
&& + \frac{1}{2} \lambda''_{ijk} (U^c)_i (D^c)_j (D^c)_k \ ,
\label{eq:superpotential}
\end{eqnarray}
with $i,j,k=1,2,3$ indicating the generation, $a,b=1,2$ the $SU(2)_L$
indices. 
The $SU(3)_c$ indices have been omitted. 
The lepton left-chiral superfields $L$ and $E^c$ are respectively $SU(2)_L$ doublet and singlet. 
The quark superfields $Q$, $U^c$ and $D^c$ denote respectively the quark $SU(2)_L$ doublet, up quark singlet and down quark singlet left-chiral superfields.
The bilinear term has been omitted in our discussion. We also neglected the soft breaking terms in the RPV sector.
Also baryon number violating RPV interactions ($\lambda''_{ijk}$) will be omitted from now, to avoid rapid proton decay.
This RPV superpotential gives the following lepton number violating Yukawa interactions:
\begin{widetext}
\begin{eqnarray}
{\cal L }_{\rm R\hspace{-.5em}/\,} &=&
- \frac{1}{2} \lambda_{ijk} \left[
\tilde \nu_i \bar e_k P_L e_j +\tilde e_{Lj} \bar e_k P_L \nu_i + \tilde e_{Rk}^\dagger \bar \nu_i^c P_L e_j -(i \leftrightarrow j ) \right] + ({\rm h.c.})\nonumber\\
&&-\lambda'_{ijk} \left[
\tilde \nu_i \bar d_k P_L d_j + \tilde d_{Lj} \bar d_k P_L \nu_i +\tilde d_{Rk}^\dagger \bar \nu_i^c P_L d_j -\tilde e_{Li} \bar d_k P_L u_j - \tilde u_{Lj} \bar d_k P_L e_i - \tilde d_{Rk}^\dagger \bar e_i^c P_L u_j \right]  + ({\rm h.c.})  \ .
\end{eqnarray}
\end{widetext}
The above fields used in our analysis are assumed to be mass-eigenstates.

\begin{figure*}[htb]
\includegraphics[width=14cm]{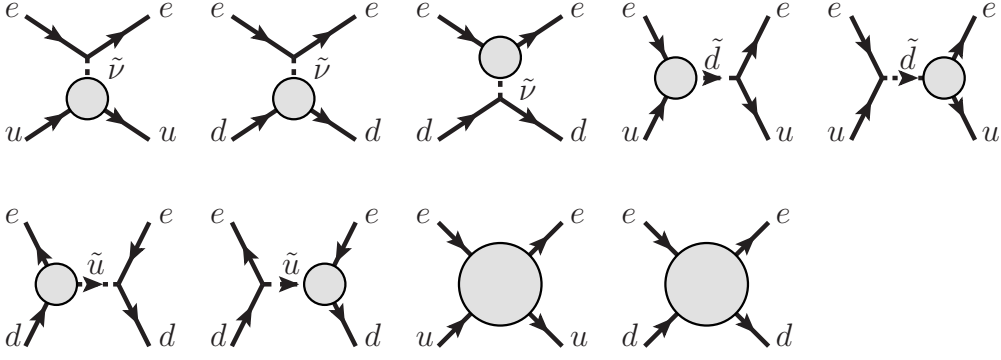}
\caption{\label{fig:classification} Classification of one-loop correction contributing to P, CP-odd $e-N$ interactions in RPVMSSM.}
\end{figure*}

\begin{figure}[htb]
\includegraphics[width=5cm]{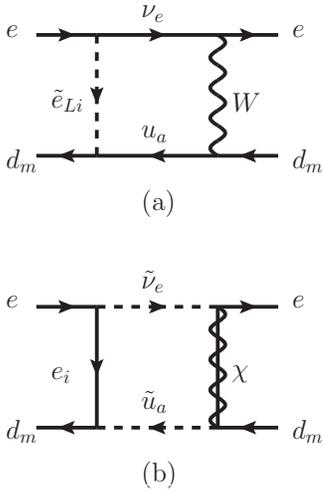}
\caption{\label{fig:diagram} Two diagrams (+ complex conjugates) contribute to the P, CP-odd $e-N$ interactions at the one-loop level in RPVMSSM.
The chargino is denoted by $\chi$.
}
\end{figure}

The possible types of one-loop correction are shown in Fig. \ref{fig:classification}.
Among the listed diagrams, the vertex corrections are renormalizations of the tree level RPV couplings, so we do not need to consider them.
This reduces our analysis only to the box diagrams.
In the evaluation of the diagrams, the Yukawa couplings of the 1st and 2nd generations are neglected.
The masses of light fermions are neglected.
We have also assumed that the soft breaking squark and slepton mass matrices have no off-diagonal components, and diagonal components do not have any CP violating phases.
For the RPV interactions, the dominance of single bilinear of RPV interactions is assumed.
With these assumptions, there are only two contributing diagrams (with their complex conjugates), which are shown in Fig. \ref{fig:diagram}.

The amplitude due to Fig. \ref{fig:diagram} (a) with its complex conjugate added is:
\begin{eqnarray}
i{\cal M}_{\rm a} & = & 8i {\rm Im} (\lambda^*_{i11} \lambda'_{iam} ) V_{am} \frac{G_F}{\sqrt{2}} m_W^2 I(m_W^2 , m_{u_a}^2 , m_{\tilde e_{Li}}^2 ) \nonumber\\
&&\ \ \ \ \ \times \left[ \bar e i\gamma_5 e \cdot \bar d_m d_m - \bar e e \cdot \bar d_m i\gamma_5 d_m  \right]  \nonumber\\
&&+\mbox{(P-even terms)} ,
\end{eqnarray}
where we have neglected the external and exchanged momenta.
We see that this amplitude is sensitive to the CP phase difference of the RPV coupling $\lambda_{i11}$ and $\lambda'_{iam}$.
$a$, $i$ and $m$ are the flavor indices.
The Fermi constant is denoted $G_F$, and $V_{am}$ is the CKM matrix element with $a$ and $m$ the flavor indices.
Here, $m_W$, $m_{u_a}$ and $m_{\tilde e_{Li}}$ are the respective masses of $W$ boson, up type quark (with flavor index $a$) and charged slepton (with flavor $i$).
The loop integral $I$ is defined as follows:
\begin{equation}
I(a,b,c) \equiv \frac{1}{4(4\pi )^2} \frac{1}{a-b} \left[ \frac{a}{c-a} \log \frac{c}{a} -\frac{b}{c-b} \log \frac{c}{b} \right] .
\end{equation}
For example, we have for $m_{\tilde e_{Li}} = 100$ GeV
\begin{eqnarray}
m_W^2 I(m_W^2 , m_u^2 , m_{\tilde e_{Li}}^2 = (100\, {\rm GeV})^2) &\approx & 1.26 \times 10^{-3} \ , \nonumber\\
m_W^2 I(m_W^2 , m_c^2 , m_{\tilde e_{Li}}^2 = (100\, {\rm GeV})^2) &\approx & 1.26  \times 10^{-3} \ , \nonumber\\
m_W^2 I(m_W^2 , m_t^2 , m_{\tilde e_{Li}}^2 = (100\, {\rm GeV})^2) &\approx & 3.7 \times 10^{-4} \ , \nonumber\\
\end{eqnarray}
and for $m_{\tilde e_{Li}} = 1$ TeV, 
\begin{eqnarray}
m_W^2 I(m_W^2 , m_u^2 , m_{\tilde e_{Li}}^2 = (1\, {\rm TeV})^2) &\approx & 5.19 \times 10^{-5} \ , \nonumber\\
m_W^2 I(m_W^2 , m_c^2 , m_{\tilde e_{Li}}^2 = (1\, {\rm TeV})^2) &\approx & 5.19 \times 10^{-5} \ , \nonumber\\
m_W^2 I(m_W^2 , m_t^2 , m_{\tilde e_{Li}}^2 = (1\, {\rm TeV})^2) &\approx & 3.30 \times 10^{-5} \ .
\end{eqnarray}
We see that this integral has a sharp dependence on the slepton mass.

The amplitude of Fig. \ref{fig:diagram} (b) (with its complex conjugate added) is:
\begin{eqnarray}
i{\cal M}_{\rm b} & = & 8i {\rm Im} (\lambda^*_{i11} \lambda'_{iam} ) V_{am} \frac{G_F}{\sqrt{2}} m_W^2 \sum_j |Z_+^{1j}|^2 \nonumber\\
&& \times I(m_{\chi_j}^2 , m_{\tilde \nu_e}^2 , m_{\tilde u_{La}}^2 ) \nonumber\\
&&\times \left[ \bar e i\gamma_5 e \cdot \bar d_m d_m - \bar e e \cdot \bar d_m i\gamma_5 d_m  \right]  . \nonumber\\
\end{eqnarray}
Here $m_{\chi_j}$, $m_{\tilde \nu_e}$ and $m_{\tilde u_{La}}$ are respectively the masses of chargino, sneutrino (1st generation) and up type squark (with flavor index $a$). 
The mixing matrix elements of the chargino $Z_+^{1j}$ ($j$=1,2) follow the notation of Rosiek \cite{rosiek}.
The important point is that ${\cal M}_{\rm a}$ and ${\cal M}_{\rm b}$ have exactly the same combinations of RPV couplings ${\rm Im} (\lambda^*_{i11} \lambda'_{iam} )$ with the same sign. 
The diagram (b) of Fig. \ref{fig:diagram} involves three sparticles in the loop, and the integral $I(m_{\chi_j}^2 , m_{\tilde \nu_e}^2 , m_{\tilde u_{La}}^2 )$ has three unknown variables.
The experimental result of the LHC has excluded the squark masses less than 1 TeV \cite{susylhc}.
We set $m_{\tilde u_{La}}=1$ TeV in this discussion.
We show some values of the integral $m_W^2 I(m_{\chi_j}^2 , m_{\tilde \nu_e}^2 , m_{\tilde u_{La}}^2 )$ for some tentative masses of sneutrino and chargino:
$m_W^2 I(m_{\chi_j}^2 , m_{\tilde \nu_e}^2 , m_{\tilde u_{La}}^2 ) = 3.8 \times 10^{-5}$ for $m_{\chi_j} = m_{\tilde \nu_e} =100$ GeV;
$m_W^2 I(m_{\chi_j}^2 , m_{\tilde \nu_e}^2 , m_{\tilde u_{La}}^2 ) = 9.9 \times 10^{-6}$ for $m_{\chi_j} = 1$ TeV and $m_{\tilde \nu_e} =100$ GeV (or for $m_{\chi_j} = 100$ GeV and $m_{\tilde \nu_e} =1$ TeV );
$m_W^2 I(m_{\chi_j}^2 , m_{\tilde \nu_e}^2 , m_{\tilde u_{La}}^2 ) = 5.1 \times 10^{-6}$ for $m_{\chi_j} = m_{\tilde \nu_e} =1$ TeV.
The order of magnitude of $m_W^2 I$ stays around $10^{-5} \sim 10^{-6}$ due to the heavy squark mass.
The relative size between ${\cal M}_{\rm a}$ depends then on the slepton mass.
We can divide the discussion in two cases.
If the slepton mass is around 100 GeV (light slepton), the amplitude ${\cal M}_{\rm a}$ has a dominant contribution.
On the other hand, when we have slepton mass $\approx 1$ TeV (heavy slepton), the amplitudes ${\cal M}_{\rm a}$ and ${\cal M}_{\rm b}$ have same order contribution.
We must note that both amplitudes have the same couplings with the same sign, so there is no possibility for cancellation with each other
The constraints on RPV couplings can therefore always be discussed.

\section{\label{sec:qcd}Derivation of the $^{199}$H\lowercase{g} EDM}
In this section, we will derive the P, CP-odd $e-N$ interactions from the one-loop RPV contributions.
The general P, CP-odd $e-N$ interactions with non-derivative couplings are
\begin{eqnarray}
H&=&\frac{G_F}{\sqrt{2}} \left[
C_N^{SP}  \bar e i\gamma_5 e \cdot \bar N N +  C_N^{PS} \bar e e \cdot \bar N i \gamma_5 N
\right.
\nonumber\\
&&\left.
\ \ \ \ \ \ \ \ \ \ \ \ \ \ \ \ \ \ \ \ \ \ +C_N^T \frac{1}{2} \epsilon^{\mu \nu \rho \sigma} \bar e \sigma_{\mu \nu } e \cdot \bar N \sigma_{\rho \sigma} N
\right] ,
\nonumber\\
\end{eqnarray} 
where $N$ denotes nucleons ($N=p$ for proton and $N=n$ for neutron).

The contribution to the scalar-pseudoscalar coupling $C_N^{SP}$ and pseudoscalar-scalar $C_N^{PS}$ coupling can be calculated by means of quark condensates in nucleons. 
\begin{eqnarray}
C_p^{SP}&=& \langle p| \bar d d |p \rangle  \tau_1 + \langle p| \bar s s |p \rangle  \tau_2 +  \langle p| \bar b b |p \rangle  \tau_3 \ , \nonumber\\
\label{eq:cnsp} C_n^{SP}&=& \langle p| \bar u u |p\rangle  \tau_1 + \langle p| \bar s s |p\rangle  \tau_2 +  \langle p| \bar b b |p \rangle  \tau_3 \ ,
\end{eqnarray}
and
\begin{eqnarray}
C_p^{PS}&=& -\langle p| \bar d i\gamma_5 d |p\rangle  \tau_1 - \langle p| \bar s i\gamma_5 s |p \rangle  \tau_2 -  \langle p| \bar b i\gamma_5 b |p \rangle  \tau_3 \ , \nonumber\\
\label{eq:cnps} C_n^{PS}&=& -\langle p| \bar u i\gamma_5 u |p  \rangle  \tau_1 - \langle p| \bar s i\gamma_5 s |p \rangle  \tau_2 -  \langle p| \bar b i\gamma_5 b |p \rangle  \tau_3 \ , \nonumber\\
\end{eqnarray}
where
\begin{eqnarray}
\tau_m & = &
-8 {\rm Im} (\lambda^*_{i11} \lambda'_{iam} ) V_{am} m_W^2 ( \  I(m_W^2 , m_{u_a}^2 , m_{\tilde e_{Li}}^2 ) 
\nonumber\\
&&\ \ \ \ \ \ \ \ \ \ \ \ \ \ \ \ \ \ \ \ +\sum_j |Z_+^{1j}|^2 I(m_{\chi_j}^2 , m_{\tilde \nu_e}^2 , m_{\tilde u_{La}}^2 ) \ ) . \nonumber\\
\end{eqnarray}
The condensates $\langle p | \bar q q | p \rangle$ and $ \langle p | \bar q i\gamma_5  q | p \rangle $ ($q=u,d,s,b$) denote respectively the scalar and pseudoscalar quark condensate in the proton.
Eqs. (\ref{eq:cnsp}) and (\ref{eq:cnps}) were derived by using isospin symmetry.

The evaluation of $\langle p| \bar q q |p \rangle$ needs non-perturbative methods of QCD calculation.
In this discussion, we use the following values for the quark contents of the proton:
\begin{eqnarray}
\langle p| \bar u u | p \rangle &=& 7.7 , \\
\langle p| \bar d d | p \rangle &=& 6.9 , \\
\langle p| \bar s s | p \rangle &=& 0.1 , \\
\langle p| \bar b b | p \rangle &=& 1 \times 10^{-2} \, .
\end{eqnarray}
The scalar condensates $\langle p| \bar u u | p \rangle$ and $\langle p| \bar d d | p \rangle$ were derived from the nucleon sigma term ($\sigma \equiv \frac{m_u + m_d }{2} \langle p | \bar u u + \bar d d| p \rangle \approx 55$ MeV \cite{gasser}) and the proton-neutron mass splitting $m_p^0 -m_n^0 = -2.05$ MeV (nucleon masses without electromagnetic contribution) \cite{zhitnitsky}, with current quark masses $m_u = 2.5$ MeV, and $m_d= 5$ MeV.
The nucleon sigma term was also evaluated in lattice QCD and gives a consistent result \cite{young}. 
The strange quark content was obtained with the recent lattice QCD calculations \cite{jlqcd,qcdsf}, with strange quark mass $m_s = 100$ MeV.
We must note that the strange quark content of nucleon evaluated in lattice QCD differs significantly from the classic result using baryon mass splitting, which gives more than one order of magnitude larger strange quark condensate.
This is due to the fact that the expansion in strange quark mass does not work accurately in the classic approach.
The bottom quark contribution was calculated using the heavy quark expansion \cite{zhitnitsky}.

For the calculation of pseudoscalar-scalar coupling $C_N^{PS}$, we need to calculate the matrix element $\langle \bar q i \gamma_5 q \rangle $ ($q=u,d,s,b$).
The calculation was done by using current algebra method with axial anomaly \cite{cheng}, which yields
\begin{equation}
\langle p| \bar q i \gamma_5 q |p \rangle =
\frac{m_N}{m_q} \left( \Delta q' + \frac{\alpha_s}{2\pi}\Delta g \right) ,
\end{equation}
where $\Delta q'$ is the fraction of the axial vector current of the quark $q$ in the proton, and $\Delta g$ is defined by $\langle p | {\rm Tr }\, G_{\mu \nu}\tilde G^{\mu \nu} | p\rangle = - 2 m_N \Delta g \bar u_p i \gamma_5 u_p$ \cite{cheng}, where $G_{\mu \nu}$ is the gluon field strength and $\tilde G_{\mu \nu}$ its dual.
We use $\Delta u' =0.82$, $\Delta d' =-0.44$, $\Delta s' =-0.11$ \cite{hermes,ellis}, $(\alpha_s / 2\pi )\Delta g = -0.16$ \cite{cheng} and recent values of quark masses cited above. This gives
\begin{eqnarray}
\langle p |  \bar u i\gamma_5 u | p \rangle &=& 248 , \\
\langle p | \bar d i\gamma_5 d | p \rangle &=& -115 , \\
\langle p | \bar s i\gamma_5 s | p \rangle &=& -2.5 , \\
\langle p | \bar b i\gamma_5 b | p \rangle &=& -3 \times10^{-2} ,
\end{eqnarray}
The pseudoscalar condensate of the bottom quark in the last line was calculated with the heavy quark expansion, in the same way as the scalar one \cite{zhitnitsky}.

The final step is the derivation of the dependence of the EDM of $^{199}$Hg atom on the P, CP-odd $e-N$ interactions.
The relativistic Hartree-Fock calculation improved with random phase approximation gives the following result \cite{flambaumdiamagnetic,flambaumformula}:
\begin{eqnarray}
d_{\rm Hg} &=& 
\left( -50 (0.40 C^{\rm SP}_p +0.60 C^{\rm SP}_n ) \right. \nonumber\\
&& \ \ \ \ \left.+ 6(0.09 C^{\rm PS}_p +0.91 C^{\rm PS}_n ) \ \right) \times 10^{-23} e\, {\rm cm} \ . \nonumber\\
\end{eqnarray}
The dependence of the $^{199}$Hg EDM on the P, CP-odd scalar-pseudoscalar interaction ($C_N^{\rm SP}$) arises via the hyperfine interaction between atomic electrons and nuclear magnetic moment \cite{flambaumformula}.
The calculation of the $^{199}$Hg atom wave function was also done within the many-body method based on relativistic coupled-cluster theory \cite{latha}, and this provides 4/3 times larger result.
This means that if we use this result, we can obtain 4/3 times tighter constraint on RPV couplings.

The final dependence of the $^{199}$Hg EDM on the supersymmetric RPV contribution can be written as
\begin{eqnarray}
d_{\rm Hg } = ( -166 \tau_1 +1.0 \tau_2 -0.032\tau_3 )\times 10^{-22} e \, {\rm cm} \ .
\label{eq:hgedmdependence}
\end{eqnarray}
Using the experimental data of $^{199}$Hg EDM (Eq. (\ref{eq:hgedmlimit})), the following limits are given
\begin{eqnarray}
|\tau_1|&<& 1.9\times 10^{-9}\ , \nonumber\\
|\tau_2|&<& 3.1\times 10^{-7}\ , \nonumber\\
|\tau_3|&<& 9.7\times 10^{-6}\ .
\end{eqnarray}

We should add a little comment on the sensitivity of the EDM of the paramagnetic $^{205}$Tl atom on $\tau_i$'s, since for equal experimental bounds, the limits on P, CP-odd scalar-pseudoscalar type interaction ($C_N^{\rm SP}$) provided by paramagnetic atomic EDM are tighter than those of diamagnetic atoms.
The dependence of $^{205}$Tl EDM on $C_N^{\rm SP}$ is \cite{flambaumtl}
\begin{eqnarray}
d_{\rm Tl} = 
-7.0  \times 10^{-18} (0.40 C^{\rm SP}_p +0.60 C^{\rm SP}_n ) \, e\, {\rm cm} \ .
\end{eqnarray}
By combining the above relation with Eq. (\ref{eq:cnsp}), we obtain
\begin{eqnarray}
d_{\rm Tl } = -(  52 \tau_1 + 0.70 \tau_2 + 0.070\tau_3 )\times 10^{-18} e \, {\rm cm} \ .
\label{eq:tledmdependence}
\end{eqnarray}
Using the experimental limit of the EDM of the $^{205}$Tl atom \cite{regan}
\begin{equation}
d_{\rm Tl} < 9.4 \times 10^{-25} e\, {\rm cm} ,
\end{equation}
we obtain
\begin{eqnarray}
|\tau_1|&<& 1.8\times 10^{-8}\ , \nonumber\\
|\tau_2|&<& 1.3\times 10^{-6}\ , \nonumber\\
|\tau_3|&<& 1.3\times 10^{-5}\ ,
\end{eqnarray}
which are looser than those given by $^{199}$Hg EDM experimental data, but do not differ by more than one order of magnitude.

\section{\label{sec:analysis}Analysis and constraints on RPV interactions from $^{199}$H\lowercase{g} EDM}

By combining Eq. (\ref{eq:hgedmdependence}) and the current experimental limit of the EDM of the $^{199}$Hg atom (see Eq. (\ref{eq:hgedmlimit}) ), we obtain the constraints on bilinears of RPV couplings as shown in Tables \ref{table:rpve-nlimitstev} and \ref{table:rpve-nlimits100gev}.
\begin{table}[htb]
\caption{Upper bounds to the RPV couplings given by the $^{199}$Hg EDM experimental data via the P, CP-odd $e-N$ interactions for $m_{\rm SUSY}=1$ TeV.
Limits from other experiments \cite{chemtob,barbier,rpvphenomenology} are also shown.
}
\begin{ruledtabular}
\begin{tabular}{ccc}
RPV couplings & $^{199}$Hg EDM & Other experiments \\ 
\hline
$|{\rm Im } (\lambda^*_{211} \lambda'_{221} )|$& $2.0 \times 10^{-5}$ &$2.9\times 10^{-2}$  \\
$|{\rm Im } (\lambda^*_{311} \lambda'_{321} )|$& $2.0 \times 10^{-5}$ & $1.7\times 10^{-2}$ \\
$|{\rm Im } (\lambda^*_{211} \lambda'_{231} )|$& $8.2 \times 10^{-4}$ & 0.60 \\
$|{\rm Im } (\lambda^*_{311} \lambda'_{331} )|$& $8.2 \times 10^{-4}$ & 0.36 \\
$|{\rm Im } (\lambda^*_{211} \lambda'_{212} )|$& $3.3 \times 10^{-3}$ &$2.9\times 10^{-2}$ \\
$|{\rm Im } (\lambda^*_{311} \lambda'_{312} )|$& $3.3 \times 10^{-3}$ &$1.7\times 10^{-2}$ \\
$|{\rm Im } (\lambda^*_{211} \lambda'_{232} )|$& $2.9 \times 10^{-2}$ &0.60 \\
$|{\rm Im } (\lambda^*_{311} \lambda'_{332} )|$& $2.9 \times 10^{-2}$ &0.36\\
$|{\rm Im } (\lambda^*_{211} \lambda'_{213} )|$& 7 & $2.9\times 10^{-2}$ \\
$|{\rm Im } (\lambda^*_{311} \lambda'_{313} )|$& 7 & $1.7\times 10^{-2}$ \\
$|{\rm Im } (\lambda^*_{211} \lambda'_{223} )|$& 0.6  & $2.9\times 10^{-2}$ \\
$|{\rm Im } (\lambda^*_{311} \lambda'_{323} )|$& 0.6  & $1.7\times 10^{-2}$ \\
\hline
\end{tabular}
\end{ruledtabular}
\label{table:rpve-nlimitstev}
\end{table}
\begin{table}[htb]
\caption{Upper bounds to the RPV couplings given by the $^{199}$Hg EDM experimental data via the P, CP-odd $e-N$ interactions for $m_{\tilde e}=100$ GeV.
Limits from other experiments \cite{chemtob,barbier,rpvphenomenology} are also shown.
}
\begin{ruledtabular}
\begin{tabular}{ccc}
RPV couplings & $^{199}$Hg EDM & Other experiments \\ 
\hline
$|{\rm Im } (\lambda^*_{211} \lambda'_{221} )|$& $8.2 \times 10^{-7}$ & $2.9\times 10^{-3}$  \\
$|{\rm Im } (\lambda^*_{311} \lambda'_{321} )|$& $8.2 \times 10^{-7}$ & $1.7\times 10^{-3}$ \\
$|{\rm Im } (\lambda^*_{211} \lambda'_{231} )|$& $7.3 \times 10^{-5}$ & $6.0\times 10^{-2}$\\
$|{\rm Im } (\lambda^*_{311} \lambda'_{331} )|$& $7.3 \times 10^{-5}$ & $3.6\times 10^{-2}$ \\
$|{\rm Im } (\lambda^*_{211} \lambda'_{212} )|$& $1.4 \times 10^{-4}$ &$2.9\times 10^{-3}$ \\
$|{\rm Im } (\lambda^*_{311} \lambda'_{312} )|$& $1.4 \times 10^{-4}$ &$1.7\times 10^{-3}$\\
$|{\rm Im } (\lambda^*_{211} \lambda'_{232} )|$& $2.6 \times 10^{-3}$ &$6.0\times 10^{-2}$\\
$|{\rm Im } (\lambda^*_{311} \lambda'_{332} )|$& $2.6 \times 10^{-3}$ &$3.6\times 10^{-2}$ \\
$|{\rm Im } (\lambda^*_{211} \lambda'_{213} )|$& 0.3 & $2.9\times 10^{-3}$  \\
$|{\rm Im } (\lambda^*_{311} \lambda'_{313} )|$& 0.3 & $1.7\times 10^{-3}$  \\
$|{\rm Im } (\lambda^*_{211} \lambda'_{223} )|$& $3 \times 10^{-2}$  & $2.9\times 10^{-3}$  \\
$|{\rm Im } (\lambda^*_{311} \lambda'_{323} )|$& $3 \times 10^{-2}$ & $1.7\times 10^{-3}$  \\
\hline
\end{tabular}
\end{ruledtabular}
\label{table:rpve-nlimits100gev}
\end{table}
In this analysis, the dominance of single bilinear of RPV couplings were assumed.
The limits given for ${\rm Im } (\lambda^*_{i11} \lambda'_{i21} )$, ${\rm Im } (\lambda^*_{i11} \lambda'_{i31} )$, ${\rm Im } (\lambda^*_{i11} \lambda'_{i12} )$ and ${\rm Im } (\lambda^*_{i11} \lambda'_{i32} )$ shows tighter constraints than limits from other experiments \cite{chemtob,barbier,rpvphenomenology}.
The new constraints were set because of the strong upper limit of the $^{199}$Hg atom EDM and also of the high sensitivity of the P, CP-odd $e-N$ interaction to the RPVMSSM.

We must pay attention to other processes contributing to the EDM of $^{199}$Hg atom induced by the same bilinears of RPV couplings discussed in our analysis.
These RPV couplings contribute to the Barr-Zee type diagram of the electron and quark EDMs with $W$ boson and charged slepton exchange.
However these contributions are small for the following reasons.
The electron EDM is suppressed  since the $^{199}$Hg atom is a diamagnetic atom.
The quark EDM generated from the RPV bilinears cited above is also negligible, since the contributing Barr-Zee type diagram has an electron loop (the Barr-Zee type contribution receives a factor of the mass of the inner loop fermion).

This analysis is expected to be also applicable to the P, CP-odd four-quark interaction at the one-loop level.
We will see below this discussion.

\section{\label{sec:4-quark}Analysis of the P, CP-odd 4-quark interaction within R-parity violation at the one-loop level}

\begin{figure}[htb]
\includegraphics[width=5cm]{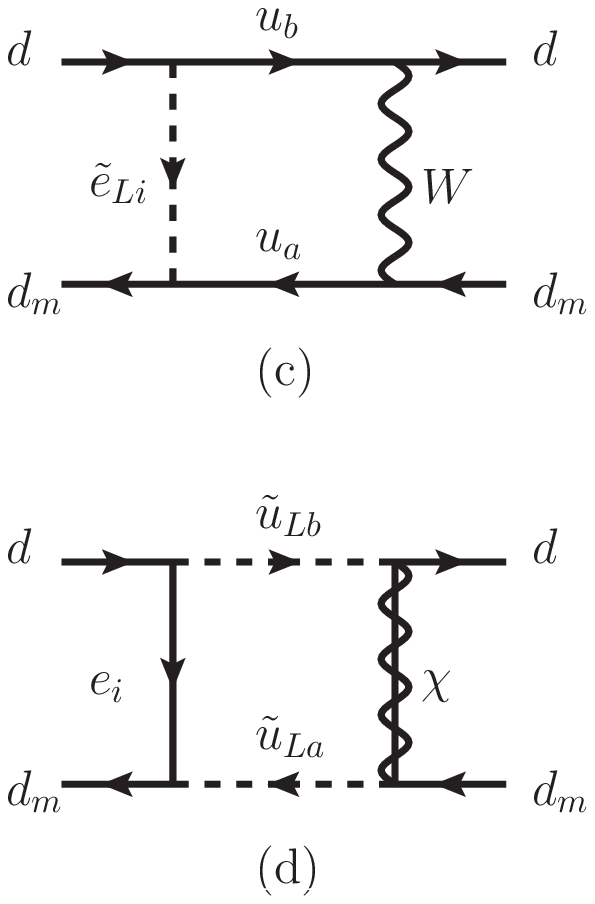}
\caption{\label{fig:4-qdiagram} RPV contribution to the P, CP-odd 4-quark interaction at the one-loop level.
The structure of the box diagram is exactly the same as the P, CP-odd electron-quark interaction.
}
\end{figure}

The previous analysis can be applied also to the P, CP-odd 4-quark interaction.
This discussion is a direct extension of the tree level analysis done by Faessler {\it et al.} \cite{faessler}.
Diagrams contributing to the P, CP-odd 4-quark interaction are shown in Fig. \ref{fig:4-qdiagram}.
The amplitudes are given as follows
\begin{eqnarray}
i{\cal M}_{\rm c} & \approx & 8i {\rm Im} (\lambda'^*_{ib1} \lambda'_{iam} ) V_{b1} V_{am} \frac{G_F}{\sqrt{2}} m_W^2 \nonumber\\
&&\ \ \ \ \cdot I'_{iab} \cdot \left[ \bar d i\gamma_5 d \cdot \bar d_m d_m - \bar d d \cdot \bar d_m i\gamma_5 d_m \right] \nonumber\\
&&+\mbox{(P-even terms)} ,\\
i{\cal M}_{\rm d} & \approx & 8i {\rm Im} (\lambda'^*_{ib1} \lambda'_{iam} ) V_{b1} V_{am} \frac{G_F}{\sqrt{2}} m_W^2 \nonumber\\
&&\ \ \ \ \cdot \sum_j |Z_+^{1j}|^2  I(m_{\chi_j}^2 ,  m_{\tilde u_{La}}^2 , m_{\tilde u_{Lb}}^2 ) \nonumber\\
&&\ \ \ \ \ \ \cdot \left[ \bar d i\gamma_5 d \cdot \bar d_m d_m - \bar d d \cdot \bar d_m i\gamma_5 d_m  \right] , \nonumber\\
\end{eqnarray}
where $i{\cal M}_{\rm c}$ and $i{\cal M}_{\rm d}$ are the amplitudes of the diagram with $W$ boson and chargino in the loop, respectively.
$a,b=1,2,3$ and $m=2,3$ are the flavor indices.
$I'_{iab}$ is the loop integral of ${\cal M}_{\rm c}$ where
\begin{equation}
I'_{iab}=
\left\{
\begin{array}{ll}
I(m_W^2 , 0 , m_{\tilde e_{Li}}^2 )  & (\mbox{a=1,2, b=1,2}) \cr
I(m_W^2 , m_t^2 , m_{\tilde e_{Li}}^2 )  & (\mbox{a=3, b=1,2}) \cr
I(m_W^2 , m_t^2 , m_{\tilde e_{Li}}^2 )  & (\mbox{a=1,2,b=3}) \cr
J(m_W^2 , m_{\tilde e_{Li}}^2 , m_t^2)  & (\mbox{a=3, b=3}) \cr
\end{array}
\right.
,
\end{equation}
with 
\begin{eqnarray}
J(a,b,c) &\equiv & \frac{1}{4(4\pi)^2} \frac{1}{a-b} \left[ \frac{a}{c-a} \left( 1 - \frac{a}{c-a} \ln \frac{c}{a} \right) \right. \nonumber\\
&&\ \ \ \ \ \ \ \ \ \ \ \ \ \ \ \ \ \ \ \left. -\frac{b}{c-b} \left( 1 - \frac{b}{c-b} \ln \frac{c}{b} \right) \right] \ . \nonumber\\
\end{eqnarray}
For example, we have
\begin{eqnarray}
m_W^2 J(m_W^2 , m_{\tilde e_{Li}}^2 =(100\, {\rm GeV})^2 , m_t^2) &\approx& 1.9 \times 10^{-4} \ , \nonumber\\
m_W^2 J(m_W^2 , m_{\tilde e_{Li}}^2 =(1\, {\rm TeV})^2 , m_t^2) &\approx& 2.6 \times 10^{-5} \ . \nonumber\\
\end{eqnarray}

Fig. \ref{fig:4-qdiagram} (d) involves two squarks in the loop, so its contribution is smaller than the amplitude ${\cal M}_{\rm c}$, since 
$m_W^2 I(m_{\chi_j}^2 , m_{\tilde u_{La}}^2 , m_{\tilde u_{Lb}}^2 ) = 9.9 \times 10^{-6}$ for $m_{\chi_j} = 100$ GeV, and
$m_W^2 I(m_{\chi_j}^2 , m_{\tilde u_{La}}^2 , m_{\tilde u_{Lb}}^2 ) = 5.1 \times 10^{-6}$ for $m_{\chi_j} = 1$ TeV 
(we have assumed $m_{\tilde u_{L}}\approx 1\,$TeV).
As we have seen for the P, CP-odd $e-N$ interaction, the amplitudes ${\cal M}_{\rm c}$ and ${\cal M}_{\rm d}$ have the same sign and combinations of couplings, so there is no possibility of cancellation with each other.

The P, CP-odd 4-quark interaction contributes to the P, CP-odd pion-nucleon interaction.
In this discussion, we use the factorization and PCAC reduction to derive the hadron level interaction.
This gives the following isovector type P, CP-odd pion-nucleon interaction
\begin{equation}
{\cal L} = \bar g_{\pi NN}^{(1)} \bar NN \pi^0 \ ,
\end{equation}
with
\begin{equation}
\bar g_{\pi NN}^{(1)} \approx -\frac{F_\pi m_\pi^2}{2m_d} \frac{G_F}{\sqrt{2}}  \sum_{m = 2,3} \rho_m \langle p | \, \bar d_m d_m | p \rangle \ ,
\end{equation}
where $F_\pi \approx 93$ MeV is the pion decay constant, and $m_\pi =140$ MeV is the pion mass.
The coefficient $\rho_m$ is defined as
\begin{eqnarray}
\rho_m &=& 8i {\rm Im} (\lambda'^*_{ib1} \lambda'_{iam} ) V_{b1} V_{am} m_W^2 \nonumber\\
&&\cdot \Bigl( I'_{iab} + \sum_j |Z_+^{1j}|^2  I(m_{\chi_j}^2 ,  m_{\tilde u_{La}}^2 , m_{\tilde u_{Lb}}^2 ) \Bigr) \ .
\end{eqnarray}
We must note that this factorization method has a large uncertainty.

The recent Schiff moment of the $^{199}$Hg nucleus was calculated with fully self-consistent mean-field treatment taking into account the deformation \cite{ban}.
The result is 
\begin{eqnarray}
S_{\rm Hg } &=& 0.007 g_{\pi NN} \bar g^{(1)}_{\pi NN} e \, {\rm fm}^3 \ ,
\label{eq:shg}
\end{eqnarray}
where $g_{\pi NN} \approx 12.9$ is the ordinary pseudoscalar coupling pion-nucleon coupling.
We have neglected contributions from the nucleon EDM, isoscalar and isotensor P, CP-odd pion-nucleon interactions.
The dependences of the $^{199}$Hg Schiff moment on isoscalar, isovector and isotensor P, CP-odd pion-nucleon interactions were calculated by Ban {\it et al.}, and they used five different codes.
The result is shown in Table \ref{table:schiffmoment}.
We use the average of the isovector dependence of the Schiff moment ($a_1$ of Table \ref{table:schiffmoment}) to give the relation (\ref{eq:shg}).
Note that this nuclear level calculation has also a large theoretical uncertainty (For some calculational method, the result is of opposite sign).

\begin{table}
\caption{Coefficients $a_i$ of the dependence of the Schiff moment on P, CP-odd pion-nucleon couplings ($S= g_{\pi NN} (a_0 \bar g^{(0)}_{\pi NN} +a_1 \bar g^{(1)}_{\pi NN}+a_2 \bar g^{(2)}_{\pi NN})$) in unit of $e$ fm$^3$. 
The labels HB and HFB stand for calculations in the Hartree-Fock and Hartree-Fock-Bogoliubov approximations, respectively.}
\begin{ruledtabular}
\begin{tabular}{lccc}
Model& $-a_0$\ \  &$-a_1$ \ \ &$a_2$ \ \  \\ 
\hline
SkM$^*$ (HFB) & 0.041 & $-$0.027 & 0.069  \\
SLy4 (HFB)& 0.013 & $-$0.006 & 0.024  \\
SLy4 (HF)& 0.013 & $-$0.006 & 0.022  \\
SV (HF)& 0.009 & $-$0.0001 & 0.016  \\
SIII (HF)& 0.012 & 0.005 & 0.016 \\
\hline
Average&0.018&$-$0.0068 &0.029\\
\end{tabular}
\end{ruledtabular}
\label{table:schiffmoment}
\end{table}

The dependence of the EDM of the $^{199}$Hg atom on the nuclear Schiff moment is given by \cite{flambaumdiamagnetic}
\begin{equation}
d_{\rm Hg} = -2.6 \times 10^{-17} \frac{S_{\rm Hg}}{e\, {\rm fm}^3} e \, {\rm cm} \ .
\end{equation}

The final form of the dependence of $^{199}$Hg EDM on the RPV P, CP-odd 4-quark contribution is
\begin{equation}
d_{\rm Hg } = (3.5\rho_2 + 0.35 \rho_3)\times 10^{-25} e\, {\rm cm}\ .
\label{eq:dhgrho}
\end{equation}
The constraints on RPV couplings obtained from the experimental data (\ref{eq:hgedmlimit}) \cite{griffith} are shown in Table \ref{table:rpv4qlimits}, where $m_{\tilde e_{Li}}$ is tentatively taken as 100 GeV and 1 TeV.

\begin{table}[htb]
\caption{Upper bounds to the RPV couplings given by the $^{199}$Hg EDM experimental data via the P, CP-odd 4-quark interactions.}
\begin{ruledtabular}
\begin{tabular}{crr}
RPV couplings & ($m_{\tilde e_{Li}}=$100 GeV) & ($m_{\tilde e_{Li}}=$1 TeV) \\ 
\hline
$|{\rm Im}( \lambda'^*_{i11} \lambda'_{i           1           2 })|$ &$4.0\times 10^{-2}$ &0.97   \\
$|{\rm Im}( \lambda'^*_{i 11} \lambda'_{i           3           2 })| $&  0.76  & 8.5\\  
$|{\rm Im}( \lambda'^*_{i 21} \lambda'_{i           1           2 })| $ & 0.17&4.2\\
$|{\rm Im}( \lambda'^*_{i 2 1} \lambda'_{i           2           2 } )|$ & $4.0\times 10^{-2}$&0.97\\
$|{\rm Im}( \lambda'^*_{i 2 1} \lambda'_{i           3           2 } )|$  & 3.3 &37\\
$|{\rm Im}( \lambda'^*_{i  3 1} \lambda'_{i           1           2 } )|$  & 15&170\\
$|{\rm Im}( \lambda'^*_{i  3 1} \lambda'_{i           2           2 }  )|$  &3.5 &40\\
$|{\rm Im}( \lambda'^*_{i 3 1} \lambda'_{i           3           2 }   )|$ &170&1200\\
$|{\rm Im}( \lambda'^*_{i 1 1} \lambda'_{i           1           3 }   )|$ &25  &630\\
$|{\rm Im}( \lambda'^*_{i 11} \lambda'_{i           2           3 }  )| $ &2.2 &53\\
$|{\rm Im}( \lambda'^*_{i  2 1} \lambda'_{i           1           3 }  )| $ &110 &2700\\
$|{\rm Im}( \lambda'^*_{i  2 1} \lambda'_{i           2           3 }  )| $ &9.4&230\\
$|{\rm Im}( \lambda'^*_{i  2 1} \lambda'_{i           3           3 }   )|$ &1.3 &15\\
$|{\rm Im}( \lambda'^*_{i  3 1} \lambda'_{i           1           3 }  )|$  &9900&$1.1\times10^5$\\
$|{\rm Im}( \lambda'^*_{i  3 1} \lambda'_{i           2           3 }  )|$  &840 &9400\\
$|{\rm Im}( \lambda'^*_{i  3 1} \lambda'_{i           3           3 }  )|$ &67&490\\
\end{tabular}
\end{ruledtabular}
\label{table:rpv4qlimits}
\end{table}
The limits obtained are looser than those obtained from other experiments \cite{barbier,chemtob,rpvphenomenology}, so it is not possible to obtain upper bounds on RPV interactions from the $^{199}$Hg EDM experimental data.
Nevertheless, this result shows the variety of RPV interactions $\lambda'_{ijk}$ accessible from the one-loop level P, CP-odd 4-quark interactions.

We should add to this discussion the possibility to constrain RPV interactions from future EDM experiments.
The first good candidate is the EDM of $^{225}$Ra atom.
The $^{225}$Ra EDM has a strong sensitivity on the P, CP-odd hadronic interactions, due to the large enhancement of the nuclear Schiff moment.
The dependence of the P, CP-odd 4-quark interaction on the EDM of $^{225}$Ra atom is \cite{dobaczewski,flambaumdiamagnetic}
\begin{equation}
d_{\rm Ra} = (-1.0 \rho_2 - 0.1 \rho_3)\times 10^{-21} e\, {\rm cm}\ .
\end{equation}
We see that the sensitivity on the P, CP-odd 4-quark interaction is enhanced by a factor of 3000 compared to the EDM of $^{199}$Hg atom (compare with eq. (\ref{eq:dhgrho})).
This large enhancement is due to the enhancement of the Schiff moment by the octupole deformation of the $^{225}$Ra nucleus \cite{dobaczewski} and the close parity doublet states of the atomic energy level \cite{flambaumdiamagnetic}.
The $^{225}$Ra EDM experiment is prepared by the group of Argonne National Laboratory aiming at the sensitivity of $O(10^{-28})e \, {\rm cm}$ \cite{mueller}.
We can then expect upper bounds of Table \ref{table:rpv4qlimits} to be tightened by several hundred times.

Another experimental candidate is the EDM of the deuteron (nucleus).
Recently a new generation of EDM experiment using the storage ring is in preparation , and it offers the possibility to measure the EDM of charged particles with very high sensitivity \cite{storage}.
With this setup, the high sensitivity to the hadronic P, CP violation is possible, since the suppression of the nuclear level P, CP violation by Schiff's  screening theorem of is avoided due to the absence of screening electrons, in addition to the long coherence time during the measurement.
The dependence of the P, CP-odd 4-quark interaction on the EDM of the deuteron is \cite{liu}
\begin{equation}
d_D = (2.9 \rho_2 + 0.29 \rho_3)\times 10^{-21} e\, {\rm cm}\ .
\label{eq:ddrho}
\end{equation}
We see that the deuteron EDM is 10000 times sensitive than the EDM of $^{199}$Hg atom against P, CP-odd 4-quark interaction.
The measurement of deuteron EDM is in preparation at Brookhaven National Laboratory, and the expected experimental sensitivity is $O(10^{-29}) e\, {\rm cm}$ \cite{bnl}.
We thus expect upper bounds of Table \ref{table:rpv4qlimits} to be tightened by $\sim$10000 times, and the deuteron EDM experiment is therefore very promising.

We should also point out the potential importance of the nucleon EDMs.
In our discussion, we have neglected the contribution of the P, CP-odd 4-quark interaction to the nucleon EDM, since the isovector type P, CP-odd pion-nucleon interaction does not contribute to the nucleon EDM in the approximation taking the leading chiral logarithm.
This approximation however neglects non-leading terms which, although being model dependent \cite{liu,dn4q}, can involve sizable isovector dependence. 
This approximation has a large theoretical uncertainty, and accurate evaluation of the dependence of nucleon EDMs on P, CP-odd quark level interactions is needed.
We are thus waiting for Lattice QCD calculation.
If we assume that the dependences of the nucleon EDM on isovector type and other P, CP-odd pion-nucleon interactions are comparable, the dependence of the P, CP-odd 4-quark interaction on the nucleon EDM will be
\begin{equation}
d_N \sim (0.1 \rho_2 + 0.01 \rho_3)\times 10^{-20} e\, {\rm cm}\ .
\end{equation}
The next generation of the neutron EDM experiments using UCN sources plans to reach the sensitivity of $O(10^{-28})e\, {\rm cm}$ \cite{ucn}, which can constrain the RPV interactions listed in Table \ref{table:rpv4qlimits} 1000 times tighter than the current $^{199}$Hg EDM experimental data.
We must also note that the proton EDM can be a strong candidate in limiting these RPV interactions, since its performance should provide the same order sensitivity as the deuteron EDM \cite{storage,bnl}.

\section{\label{sec:conclusion}Conclusion}
In this discussion we have analyzed the contribution of the RPVMSSM to the P, CP-odd $e-N$ interaction at the one-loop level and have derived from the recent $^{199}$Hg EDM experimental data limits to the imaginary parts of the following products of RPV couplings: $\lambda^*_{i11} \lambda'_{i21}$, $\lambda^*_{i11} \lambda'_{i31}$, $\lambda^*_{i11} \lambda'_{i12}$, $\lambda^*_{i11} \lambda'_{i32}$, $\lambda^*_{i11} \lambda'_{i13}$ and $\lambda^*_{i11} \lambda'_{i23}$ ($i=2,3$).
For $\lambda^*_{i11} \lambda'_{i21}$, $\lambda^*_{i11} \lambda'_{i31}$, $\lambda^*_{i11} \lambda'_{i12}$ and $\lambda^*_{i11} \lambda'_{i32}$ ($i=2,3$), we have found that these limits give tighter constraints than those given by other experiments. For $\lambda^*_{i11} \lambda'_{i13}$ and $\lambda^*_{i11} \lambda'_{i23}$ ($i=2,3$), we could not set new limits.
The new constraints were set because of the strong upper limit of the $^{199}$Hg atom EDM and also of the high sensitivity of P, CP-odd $e-N$ interaction to the RPVMSSM.
We have also analyzed the P, CP-odd 4-quark interaction within RPVMSSM at the one-loop level.
The current experimental limit of the $^{199}$Hg EDM could not set new limits, but new generation of EDM experiments with $^{225}$Ra atom, deuteron, neutron and proton has the possibility to constrain the RPV interactions significantly via P, CP-odd 4-quark interaction.
These EDM experiments are therefore very promising.

The result of our analysis has demonstrated the importance of the subleading order analysis for the EDM, like the well-known analysis of the muon anomalous magnetic moment.
It has also emphasized the accessibility to a variety of RPV interactions through the subleading loop level contributions, within the assumption of the dominance of single bilinear of RPV couplings.
We have been able to set new limits to RPV interactions thanks to the combination of the high accuracy of the EDM experimental data and the variety of RPV interactions relevant at the one-loop level.

We should also note that the limits to RPV interactions given by the analysis of the P, CP-odd 4-quark interactions has a large theoretical uncertainty and model dependence in QCD calculation.
To give more determinate constraints, accurate calculations are indispensable.
The study of the dependence of P, CP-odd hadron level interactions on P, CP-odd 4-quark interactions within Lattice QCD is therefore required.

\begin{acknowledgments}
The author thanks T. Sato and T. Kubota for useful discussions and comments.
\end{acknowledgments}

\end{document}